\documentclass[preprint]{jpsj2}
\usepackage{graphicx}

\def\runtitle{Enhancement of Kondo effect in multilevel quantum dots}
\def\runauthor{M.\ {\sc Eto}}

\title{%
Enhancement of Kondo effect in multilevel quantum dots
}

\author{%
Mikio {\sc Eto}\thanks{E-mail address: eto@rk.phys.keio.ac.jp}
}

\inst{%
Faculty of Science and Technology, Keio University,
3-14-1 Hiyoshi, Kohoku-ku, Yokohama 223-8522, Japan
}

\recdate{July 17, 2004}

\abst{%
We theoretically study enhancement mechanisms of the Kondo effect in
multilevel quantum dots. In quantum dots fabricated
on semiconductors, the energy difference between
discrete levels $\Delta$ is tunable by applying a magnetic field.
With two orbitals and spin $1/2$ in the quantum dots, we evaluate the Kondo
temperature $T_{\rm K}$ as a function of $\Delta$,
using the scaling method. $T_{\rm K}$ is maximal around $\Delta=0$
and decreases with increasing $|\Delta|$, following a power law,
$T_{\rm K}(\Delta)=T_{\rm K}(0)\cdot ( T_{\rm K}(0)/|\Delta| )^{\gamma}$,
which is understood as a crossover from SU(4) to SU(2) Kondo effect.
The exponents on both sides of a level crossing, $\gamma_{\rm L}$
and $\gamma_{\rm R}$, satisfy a relation of
$\gamma_{\rm L}\cdot\gamma_{\rm R}=1$.
We compare this enhanced Kondo effect with that by
spin-singlet-triplet degeneracy for an even number of electrons,
to explain recent experimental results using vertical quantum dots.
}
\kword{quantum dot, Kondo effect, multilevel, scaling method}

\begin{document}
\sloppy
\maketitle

\section{Introduction}

Kondo effect is one of the most important many-body problems in solid-state
physics.\cite{Kondo}
When a localized spin ($S=1/2$) is brought in contact with electron Fermi sea,
the Kondo effect gives rise to a new many-body ground state of a spin singlet
and markedly influences the transport properties.\cite{Hewson,Yosida}
The Kondo effect was originally discovered in metals including dilute
magnetic impurities.
The Kondo effect also takes place in semiconductor quantum dots connected to
external leads through tunnel barriers, as predicted by several theoretical
groups.\cite{Glazman,Ng,Kawabata,Hershfield,Hershfield2,Meir}
In this case, a localized spin $S$ is formed by electrons confined
in the quantum dots due to the Coulomb blockade.\cite{Leo}
Usually, the spin is zero or $1/2$ in the Coulomb blockade region with
an even and odd number of electrons, respectively, since discrete
spin-degenerate levels are consecutively occupied in the quantum dots.
Hence the Kondo effect is observed in the latter case
only.\cite{GGNat,Cronen,GGPRL}

The semiconductor quantum dots offer the following merits for the
investigation of the Kondo physics.
First, we can examine the properties of single ``magnetic impurity.''
Second, the transport of conduction electrons is observed through the magnetic
impurity in the case of quantum dots, whereas the scattering by the magnetic
impurities is observed in the traditional metallic systems. In the former,
the resonant nature of the Kondo effect is directly detected:
The conductance is enhanced to a universal value of $2e^2/h$ at
temperatures much below the Kondo temperature, $T_{\rm K}$,\cite{Wilfred}
which is ascribable to the resonant tunneling through Kondo singlet state.
In the latter, the electric resistance is anomalously enhanced at low
temperatures since the scattering probability is amplified by the resonance.
This explains a
``resistance minimum'' as a function of temperature ($T$), together with the
resistivity due to the phonons which increases with $T$.\cite{Kondo}
(In the Coulomb blockade region of quantum dots, the normal conductance
increases with $T$ by the thermal activation. Thus the Kondo effect results
in a ``conductance minimum.'')
Third, various parameters are tunable in quantum dots using gate electrodes,
{\it e.g}., number of electrons, strength of tunnel coupling.
Nonequilibrium situations are also realized by applying a finite bias voltage.

Besides the conventional Kondo effect with spin $1/2$,
unconventional Kondo effects have been discovered which stem from
the interplay between spin and orbital degrees of freedom in quantum dots.
An example is the enhanced Kondo effect at the energy degeneracy
between spin-singlet and -triplet states for an even number of
electrons (S-T Kondo effect).\cite{Sasaki}
In vertically fabricated quantum dots, the spacing between discrete
levels can be tuned by applying a magnetic field perpendicularly to
the quantum dots.
Around a level degeneracy, high-spin states appear by the exchange
interaction (Hund's rule).\cite{Tarucha}
Hence the energy difference $\Delta$ between spin-singlet and -triplet
states is tunable by the magnetic field, as schematically shown in Fig.\ 1(b).
The Kondo effect is significantly enhanced around $\Delta=0$.
Note that the Zeeman effect is irrelevant in the experiment
since the Zeeman energy is much
smaller than $T_{\rm K}$, owing to a small $g$ factor in semiconductors.
This S-T Kondo effect has been explained theoretically by the author and
Nazarov\cite{Eto,Eto2,Eto3} and
by Pustilnik and Glazman.\cite{Pustilnik,Pustilnik2}
(The Zeeman effect is relevant to the S-T Kondo effect in carbon
nanotubes.\cite{nanotube} We refer to ref.\ 23 
for the theoretical study.)

Recently another large Kondo effect has been found around the
two-orbital degeneracy for an odd number of electrons with spin $1/2$
[doublet-doublet (D-D) Kondo effect].\cite{Sasaki2}
When the energy difference between the two orbitals $\Delta$ is tuned
[Fig.\ 1(a)], the conductance is largely enhanced around $\Delta = 0$ by
the Kondo effect. Though the Kondo effect is expected with single level
and spin $1/2$, the Kondo temperature is probably smaller than actual
temperature $T$ so that no conductance increase is seen except the vicinity of
the level crossing point in this experiment. Around $\Delta = 0$,
the estimated $T_{\rm K}$ is similar to that for the S-T Kondo effect
observed in the same samples, indicating that a total of fourfold spin and
orbital degeneracy contributes to enhance $T_{\rm K}$ in both the cases.

The purpose of the present paper is to study the D-D Kondo effect
theoretically and to elucidate the differences between D-D and
S-T Kondo effects. It is known that the orbital degeneracy plays an important
role in the Kondo effect of magnetic impurities with $f$ electrons.
When the total degeneracy factor is $N_{\rm d}=2j+1$ ($j$ is the total
angular momentum consisting of orbital angular momentum and spin),
the Kondo effect is described by the Coqblin-Schrieffer model of
SU($N_{\rm d}$) symmetry.\cite{Coqblin,Hewson}
This model yields the Kondo temperature
of $T_{\rm K} \sim D_0 e^{-1/N_{\rm d} \nu J}$ with exchange coupling $J$,
where $D_0$ and $\nu$ are the bandwidth and density of states of
conduction electrons, respectively.
Hence the degeneracy factor $N_{\rm d}$ significantly
enhances the Kondo temperature.
We examine the Kondo effect in quantum dots with two orbitals and spin
$1/2$ and evaluate the Kondo temperature as a function of $\Delta$.
We find that $T_{\rm K}$ is maximal around $\Delta=0$ and decreases with
increasing $|\Delta|$ by a power law,
$T_{\rm K}(\Delta)=T_{\rm K}(0)\cdot ( T_{\rm K}(0)/|\Delta| )^{\gamma}$,
with exponent $\gamma=1$. This behavior of $T_{\rm K}$ is
understood as a crossover from SU(4) to SU(2) Kondo effect when
the strength of the tunnel coupling is equivalent for the two orbitals.
Since the spin-orbit interaction can be disregarded in the
quantum dots, the degenerate factor is $N_{\rm d}=4$ ($N_{\rm d}=2$) for
orbital and spin degrees of freedom when $\Delta=0$ ($\Delta \ne 0$).
In general situations, the exponent of the power law is not a universal
value since the fixed point of SU(4) Kondo effect is marginal. We find a
relation of $\gamma_{\rm L}\cdot\gamma_{\rm R}=1$, where $\gamma_{\rm L}$
and $\gamma_{\rm R}$ are the exponents on both sides of a level crossing.
The $\Delta$ dependence of $T_{\rm K}$ was not studied in details by
previous theoretical works on the Kondo effect in
multilevel quantum dots.\cite{Inoshita,Inoshita2,Pohjola,Izumida,Yeyati}

The behavior of $T_{\rm K}(\Delta)$ in the D-D Kondo effect is similar to
that in the S-T Kondo effect.\cite{Eto,Eto2,Eto3,Pustilnik,Pustilnik2}
We clarify the differences between them to explain the above-mentioned
experimental results.\cite{Sasaki2}.
Our discussion is restricted to the case of
quantum dots vertically fabricated from double-barrier
heterostructures.\cite{Sasaki,Sasaki2,Tarucha,Leo}
In transport processes between a quantum dot and leads through
heterostructure tunnel junctions, the orbital quantum numbers are conserved
for the motion of electrons in the transverse direction.
Therefore there are two channels of conduction electrons in the leads
when two orbitals are relevant in the quantum dot; each channel couples to
only one of the two orbitals.
The situation is different in quantum dots of lateral geometry, in which a
single channel in each lead couples to both the orbitals through a tunnel
junction.\cite{Leo}
We refer to refs.\ 31 
and 32--34            
for experimental and theoretical studies on the S-T Kondo effect
in lateral quantum dots.

The organization of the present paper is as follows.
In the next section (\S 2), we model a quantum dot with
two orbitals and spin $1/2$ for the D-D Kondo effect.
In \S 3, the Kondo temperature is evaluated as a function of energy
difference $\Delta$ between the two orbitals, using the scaling method.
In \S 4, the scaling analysis is given for the S-T Kondo effect, according
to our previous works.\cite{Eto,Eto2,Eto3} The final section (\S 5) is
devoted to conclusions and discussion. We compare the D-D and S-T Kondo
effects to explain the experimental results.\cite{Sasaki2}

\begin{figure}[t]
\begin{center}
\includegraphics[width=8cm]{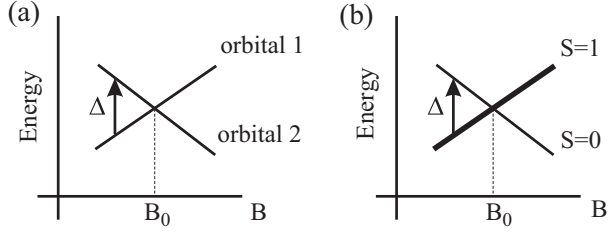}
\end{center}
\caption{In a quantum dot with two orbitals, the energy difference $\Delta$
can be tuned using a magnetic field $B$.
(a) $\Delta=\varepsilon_2-\varepsilon_1$, level spacing between the
two orbitals for an electron in the quantum dot, and
(b) $\Delta=E_{S=0}-E_{S=1}$, the energy difference between spin-singlet
and -triplet states for two electrons.}
\end{figure}

\section{Model}

As a model for the D-D Kondo effect, we consider a quantum dot with an
electron ($S=1/2$) and two orbitals ($i=1,2$). The energy difference between
the orbitals is denoted by $\Delta=\varepsilon_{2}-\varepsilon_{1}$.
The Hamiltonian in the quantum dot reads
\begin{equation}
H_{\rm{dot}} = \sum_{\mu}\varepsilon_i d_{\mu}^{\dagger} d_{\mu}
+E_{\rm c}(\hat{n}-{\cal N}_{\rm{gate}})^2,
\label{eq:Hdot}
\end{equation}
where $d_{\mu}^{\dagger}$ and $d_{\mu}$ are creation and annihilation
operators of the dot state $\mu=(i,S_z)$, respectively.
$\hat{n}=\sum_{\mu} d_{\mu}^{\dagger}d_{\mu}$ is the number operator of
electrons in the quantum dot. $E_{\rm c}$ is the charging energy,
whereas ${\cal N}_{\rm{gate}}$ is tuned by the gate voltage.
We set ${\cal N}_{\rm{gate}} = 1$ for the Coulomb blockade region with
an electron in the quantum dot.
The exchange interaction between electrons in the
quantum dot is neglected in sections 2 and 3 since it is irrelevant to the
D-D Kondo effect. 

The quantum dot is connected to two external leads, $L$ and $R$,
through tunnel barriers.
We assume that the orbital symmetry is conserved in the tunnel
processes, which is the case of vertical quantum dots.
Hence the leads have two channels;
channel $i$ ($=1,2$) in lead $L$ ($R$) couples to orbital $i$ by $V_{L,i}$
($V_{R,i}$). We perform a unitary transformation for conduction
electrons in the two leads, along the lines of ref.\ 4;  
\begin{eqnarray*}
c_{k \mu} & = & (V_{L,i}^*c_{L,k \mu}+V_{R,i}^*c_{R,k \mu})/V_i,
\\
\bar{c}_{k \mu} & = & (-V_{R,i}c_{L,k \mu}+V_{L,i}c_{R,k \mu})/V_i,
\end{eqnarray*}
with $V_i=\sqrt{|V_{L,i}|^2+|V_{R,i}|^2}$, where $c_{\alpha,k\mu}$
is the annihilation operator of an electron in lead $\alpha$,
with momentum $k$ and $\mu=(i,S_z)$.
The mode $c_{k \mu}$ is coupled to the quantum dot with $V_i$, whereas the mode
$\bar{c}_{k \mu}$ is decoupled and thus shall be disregarded.
The Hamiltonian of the leads and that of tunnel processes are written as
\begin{eqnarray}
H_{\rm{leads}} & = & \sum_{k} \sum_{\mu}
\varepsilon_{k} c_{k\mu}^{\dagger} c_{k\mu}, \\
H_{\rm{T}} & = & \sum_{k} \sum_{\mu}
V_i (c_{k\mu}^{\dagger} d_{\mu} + {\rm h.c.}),
\label{eq:tunnel}
\end{eqnarray}
respectively.

We define addition and extraction energies as the energy cost to add or
subtract an electron on/from the quantum dot,
$E^+=\varepsilon_1 + E_{\rm c}-E_{\rm F}$, $E^-=E_{\rm F}-\varepsilon_1$,
where $E_{\rm F}$ is the Fermi level in the leads.
We assume that $E^{\pm} \gg |\Delta|$.
The level broadening of orbital $i$ is given by $\Gamma_i=\pi \nu V_i^2$,
where $\nu$ is the density of states in the leads.
The level broadening and temperature $T$ should be much smaller than
$E^{\pm}$ for the Coulomb blockade.

\begin{figure}[t]
\begin{center}
\includegraphics[width=4.5cm]{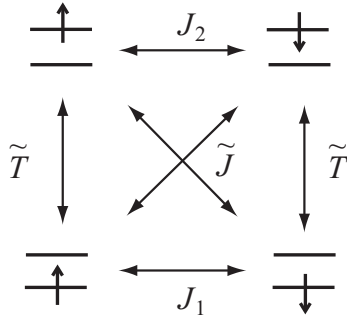}
\end{center}
\caption{Four states in a quantum dot with an electron ($S=1/2$) and
two orbitals ($i=1,2$). There are exchange processes among them ($J_1$,
$J_2$, $\tilde{T}$ and $T$) due to the second-order tunnel couplings with
conduction electrons in the leads.}
\end{figure}

Considering the second-order tunnel processes, $H_{\rm{T}}$,
we obtain the effective Hamiltonian $H_{\rm eff}$ in a subspace with
one electron in the quantum dot. We introduce the pseudofermion operators
$f^{\dagger}_{\mu}$ ($f_{\mu}$) that create (annihilate) the dot state
$\mu=(i,S_z)$. The constraint of
\begin{equation}
\sum_{\mu} f^{\dagger}_{\mu} f_{\mu}=1
\end{equation}
is required. The Hamiltonian is written as
\begin{eqnarray}
H_{\rm{eff}} & = & \sum_{\mu}\varepsilon_i f_{\mu}^{\dagger} f_{\mu}
+ H_{\rm{lead}} \nonumber \\
& + & 
2J_1 {\mbi s}^{(1)} \cdot {\mbi S}^{(1)}
+2J_2 {\mbi s}^{(2)} \cdot {\mbi S}^{(2)} \nonumber \\
& + & 2\tilde{J} \left[ {\mbi s}^{(12)} \cdot {\mbi S}^{(21)}+
            {\mbi s}^{(21)} \cdot {\mbi S}^{(12)} \right]  \nonumber \\
& + & 2\tilde{T} \left[ \left(\sum_{kk'}\sum_{s}
c_{k',1,s}^{\dagger} c_{k,2,s} \right)
\left( \sum_{s} f_{2,s}^{\dagger}f_{1,s} \right)
+{\rm h.c.} \right]  \nonumber \\
& + & 2T \left[ \sum_{kk'}\sum_{s} (c_{k',1,s}^{\dagger} c_{k,1,s}-
c_{k',2,s}^{\dagger} c_{k,2,s}) \right]
\left[ \sum_{s} (f_{1,s}^{\dagger}f_{1,s}- f_{2,s}^{\dagger}f_{2,s})
\right],
\label{eq:Hmltn0}
\end{eqnarray}
where
\begin{eqnarray*}
{\mbi s}^{(i)} & = & \sum_{kk'}\sum_{ss'}
c_{k',i,s'}^{\dagger} ({\mbi \sigma}_{s's}/2) c_{k,i,s}, \\
{\mbi S}^{(i)} & = & \sum_{ss'}
f_{i,s'}^{\dagger} ({\mbi \sigma}_{s's}/2) f_{i,s}, \\
{\mbi s}^{(ij)} & = & \sum_{kk'}\sum_{ss'}
c_{k',i,s'}^{\dagger} ({\mbi \sigma}_{s's}/2) c_{k,j,s}, \\
{\mbi S}^{(ij)} & = & \sum_{ss'}
f_{i,s'}^{\dagger} ({\mbi \sigma}_{s's}/2) f_{j,s}.
\end{eqnarray*}
The third (fourth) term in $H_{\rm{eff}}$ represents the exchange coupling
between conduction electrons of channel 1 (2) and dot spin $1/2$ at orbital
1 (2). The terms with $\tilde{J}$ and $\tilde{T}$ describe the interchannel
scattering of conduction electrons by the dot electron
with or without spin flip. The coupling constants are given by
$J_1=V_1^2/\tilde{E}_{\rm C}$, $J_2=V_2^2/\tilde{E}_{\rm C}$
and $\tilde{J}=\tilde{T}=V_1V_2/\tilde{E}_{\rm C}$, where
$1/\tilde{E}_{\rm C}=1/E^++1/E^-$.
These processes are schematically shown in Fig.\ 2.
A potential scattering term with
$T=(V_1^2+V_2^2)/(2\tilde{E}_{\rm C})$ is also relevant to the Kondo
effect, whereas irrelevant potential scatterings are
disregarded in $H_{\rm{eff}}$.

In the case of $V_1=V_2$, all the coupling constants are identical to
one another, $J_1=J_2=\tilde{J}=\tilde{T}=T \equiv J$. Then
the effective Hamiltonian is reduced to the Coqblin-Schrieffer model of
SU(4) symmetry if $\Delta=0$.
Indeed, using a pseudospin $T_z= \pm 1/2$ to represent orbitals $i=1,2$,
$H_{\rm eff}$ is rewritten as
\begin{equation}
H_{\rm eff}=H_{\rm{lead}}
+J {\mbi S} \cdot {\mbi s} - \tilde{\mbi B} \cdot {\mbi S},
\label{eq:Hmltn}
\end{equation}
where
\begin{eqnarray*}
{\mbi S} & = & \frac{1}{2}\sum_{\mu \nu} f_{\mu}^{\dagger}
\hat{\mbi \Sigma}_{\mu \nu} f_{\nu}, \\
{\mbi s} & = & \frac{1}{2}
\sum_{k k'}\sum_{\mu \nu} c_{k',\mu}^{\dagger}
\hat{\mbi \Sigma}_{\mu \nu} c_{k,\nu}.
\end{eqnarray*}
$\hat{\mbi \Sigma}$ involves 15 components which are
generators of the Lie algebra su(4),
$\{ (\tau_x, \tau_y, \tau_z, I) \otimes
(\sigma_x, \sigma_y, \sigma_z, I) \} - \{ I \otimes I \}$, where $\sigma_i$
($\tau_i$) are the Pauli matrices in the spin (pseudospin) space
and $I$ is the unit matrix.
The energy difference $\Delta$ is expressed by the fictitious magnetic
field $\tilde{\mbi B}$, the $\tau_z \otimes I$ component of which is
$\Delta$ and the other components are 0:
\[
\tilde{\mbi B} \cdot {\mbi S}=\frac{\Delta}{2}\sum_{s}
(f_{1,s}^{\dagger}f_{1,s}-f_{2,s}^{\dagger}f_{2,s}).
\]
When $\tilde{\mbi B}=0$, the Hamiltonian (\ref{eq:Hmltn})
is invariant with respect to the rotation in a product space of spin and
pseudospin spaces [SU(4) symmetry].

\section{Scaling calculations}

To evaluate the Kondo temperature, the effective Hamiltonian,
(\ref{eq:Hmltn0}) or (\ref{eq:Hmltn}), is analyzed using ``poor man's''
scaling method.\cite{Anderson,Nozieres,Cox}
We assume that the density of states of conduction
electrons in the leads is constant $\nu$ in the energy range $[-D,D]$.
With changing the bandwidth from $D$ to $D-|d D|$, we renormalize the
exchange coupling constants ($J_1$, $J_2$, $\tilde{J}$, $\tilde{T}$, $T$)
not to change the low-energy physics, within the second-order perturbation with
respect to the exchange couplings. This procedure yields the scaling equations
in two limits, $D \gg |\Delta|$ and $D \ll |\Delta|$.
The Kondo temperature $T_{\rm K}$ is determined as the
energy scale $D$ at which the coupling constants become so large that
the perturbation breaks down.

\subsection{SU(4) Kondo effect}

We begin with the case of $V_1=V_2$, where $J_1=J_2=\tilde{J}=\tilde{T}=T
\equiv J$. Using the Hamiltonian (\ref{eq:Hmltn}),
we obtain the scaling equations for a single coupling constant $J$.

When the energy scale $D$ is much larger than the energy difference
$|\Delta|$, the scaling equation is given by
\begin{equation}
dJ/d \ln D= -4\nu J^2.
\label{eq:scalDDA}
\end{equation}
The exchange coupling $J$ develops rapidly with decreasing $D$,
reflecting a fourfold degeneracy of spin and pseudospin.
This represents the SU(4) Kondo effect.

When $D \ll |\Delta|$, the equation is
\begin{equation}
dJ/d \ln D= -2\nu J^2.
\label{eq:scalDDB}
\end{equation}
The evolution of $J$ is slower since one component of $T_z$
(orbital of the upper level) is irrelevant.
This is the conventional Kondo effect of SU(2) symmetry.

From these equations, we evaluate the Kondo temperature $T_{\rm K}$ as a
function of $\Delta$. (i) $T_{\rm K}$ is maximal around $\Delta = 0$
[$|\Delta| \ll T_{\rm K}(\Delta=0)$].
In this case, the scaling equation (\ref{eq:scalDDA}) remains valid till
the scaling ends. This yields 
\begin{equation}
T_{\rm K}(\Delta=0)=D_0 \exp[-1/4\nu J],
\end{equation}
where $D_0$ is the initial bandwidth and given by
$\displaystyle D_0=\sqrt{E_+E_-}$.\cite{Haldane}

(ii) $T_{\rm K}$ is minimal when $|\Delta| \gg D_0$. Then eq.\
(\ref{eq:scalDDB}) is valid in the whole scaling region. Solving the
equation, we obtain
\begin{equation}
T_{\rm K}(\Delta=\infty)=D_0 \exp[-1/2\nu J].
\end{equation}

(iii) The intermediate region of $T_{\rm K}(0) \ll |\Delta| \ll D_0$ is
examined as follows. With decreasing the energy scale $D$,
the renormalization flow goes toward the fixed point of SU(4) Kondo effect
first [eq.\ (\ref{eq:scalDDA}) for $D \gg |\Delta|$],
and then goes to the fixed point of SU(2) Kondo effect
[eq.\ (\ref{eq:scalDDB}) for $D \ll |\Delta|$].
To consider this crossover behavior, we match the solutions of
eqs.\ (\ref{eq:scalDDA}) and (\ref{eq:scalDDB}) at
$D \approx |\Delta|$. We find that $T_{\rm K}(\Delta)$
decreases with increasing $|\Delta|$, following a power law,
\begin{equation}
T_{\rm K}(\Delta)=T_{\rm K}(0)\cdot
( T_{\rm K}(0)/|\Delta| )^{\gamma},
\label{eq:powerDD}
\end{equation}
with exponent $\gamma=1$.
Our results of (i), (ii) and (iii) clearly indicate an enhancement of the Kondo
effect around $\Delta=0$ [see Fig.\ 6(a) for the schematic drawing of
$T_{\rm K}(\Delta)$], in accordance with
the experimental observation.\cite{Sasaki2}

The expression of $T_{\rm K}(\Delta)$, eq.\ (\ref{eq:powerDD}),
has been originally derived by Yamada {\it et al}.\ for the Kondo effect in
metals with magnetic impurities in the presence of a crystal
field.\cite{Yamada} The exact solution of this Kondo effect has also been
obtained.\cite{Schlottmann,Kawakami}
Lately Borda {\it et al}.\ has examined the SU(4) Kondo effect in a double
quantum dot system with interdot capacitive coupling.\cite{Borda}
Their model is equivalent to ours when the tunnel coupling is absent
between the double dots. [The magnetic field in ref.\ 42  
corresponds to $\tilde{{\mbi B}}$ in eq.\ (\ref{eq:Hmltn}).]
Using numerical renormalization group calculations,
they have estimated $T_{\rm K}$ as a function of magnetic field
and shown its enhancement around the fourfold degeneracy.
This is in qualitative agreement with our scaling results.

\subsection{General case of $V_1 \ne V_2$}

Now we examine general situations of $V_1 \ne V_2$. We set
$\Delta=\varepsilon_{2}-\varepsilon_{1} \ge 0$ in this subsection.
We derive a set of scaling equations for $J_1$, $J_2$, $\tilde{J}$,
$\tilde{T}$ and $T$ in Hamiltonian (\ref{eq:Hmltn0}), using the poor man's
scaling method.
For $D \gg \Delta$,
   \begin{equation}
   \left\{
   \begin{array}{rcl}
   dJ_1/\nu d \ln D &=& -2J_1^2-\tilde{J}(\tilde{J}+\tilde{T}), \\
   dJ_2/\nu d \ln D &=& -2J_2^2-\tilde{J}(\tilde{J}+\tilde{T}), \\
   d\tilde{J}/\nu d \ln D &=& -\tilde{J}(J_1+J_2+T)-\tilde{T}(J_1+J_2)/2, \\
   d\tilde{T}/\nu d \ln D &=& -3\tilde{J}(J_1+J_2)/2-\tilde{T}T, \\
   dT/\nu d \ln D &=& -3\tilde{J}^2-\tilde{T}^2.
   \end{array}
   \right.
   \label{eq:DDscale}
   \end{equation}
For $D \ll \Delta$,
   \begin{equation}
   dJ_1/ d \ln D= -2 \nu J_1^2,
   \label{eq:DDscale2}
   \end{equation}
whereas the other coupling constants do not evolve.

Analyzing eqs.\ (\ref{eq:DDscale}), we find that the fixed point of SU(4)
Kondo effect, $J_1=J_2=\tilde{J}=\tilde{T}=T=\infty$, is marginal (Appendix A).
With a decrease in $D$, a marginal variable, $(J_1-J_2)/(J_1+J_2)$, is
almost kept at the initial value while the other variables are renormalized
toward the values at the fixed point.
In consequence, $T_{\rm K}(\Delta)$ is not a universal function although
$T_{\rm K}$ is always maximal around $\Delta=0$.

\begin{figure}[t]
\begin{center}
\includegraphics[width=6cm]{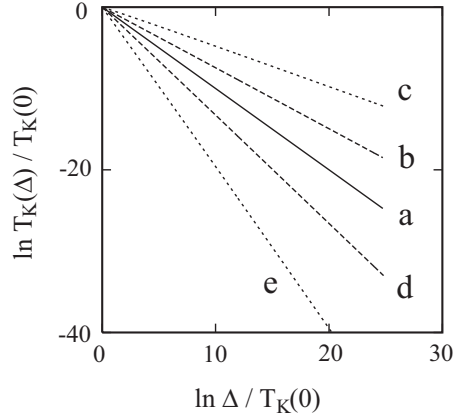}
\end{center}
\caption{
Calculated results of the Kondo temperature $T_{\rm K}$ for the D-D Kondo
effect, as a function of energy difference $\Delta$ between the two
orbitals, on a log-log scale.
Both $T_{\rm K}$ and $\Delta$ are normalized by the Kondo temperature at
$\Delta=0$, $T_{\rm K}(0)$.
(a) $(V_2/V_1)^2=1$, (b) 3/4, (c) 1/2, (d) 4/3 and (e) 2.
The scaling equations, (\ref{eq:DDscale}) for $D>\Delta$ and
(\ref{eq:DDscale2}) for $D<\Delta$, are solved numerically
with the initial condition of $\nu(J_1+J_2) = 0.02$.}
\end{figure}

To examine $T_{\rm K}(\Delta)$, we numerically solve the scaling equations,
eqs.\ (\ref{eq:DDscale}) for $D>\Delta$ and
eq.\ (\ref{eq:DDscale2}) for $D<\Delta$.
We put $\nu (J_1+J_2) = 0.02$ in the initial condition and determine
the Kondo temperature $T_{\rm K}$ as the energy scale $D$ at which
$\nu J_1$ becomes as large as unity.
The calculated results of $T_{\rm K}(\Delta)$ are shown
in Fig.\ 3 on a log-log scale, with (a) $(V_2/V_1)^2=1$,
(b) 3/4, (c) 1/2, (d) 4/3 and (e) 2. $T_{\rm K}(\Delta)$ seems to follow
a power law
\begin{equation}
T_{\rm K}(\Delta)=T_{\rm K}(0)\cdot
( T_{\rm K}(0)/\Delta )^{\gamma},
\label{eq:powerDD2}
\end{equation}
but the exponent depends on model parameters. It is approximately
given by
\begin{equation}
\gamma=\left( \frac{V_2}{V_1} \right)^2=\frac{\Gamma_2}{\Gamma_1},
\label{eq:DDgamma}
\end{equation}
as long as $\Gamma_1 \sim \Gamma_2$ (Appendix A).
If $\Gamma_2/\Gamma_1 \gg 1$ or $\ll 1$,
$T_{\rm K}(\Delta)$ deviates from the power law. Then the renormalization
of $\Delta$,\cite{Kuzmenko} that is not considered in our calculations,
might have to be taken into account to determine $T_{\rm K}$, besides the
renormalization of exchange couplings.

If we denote the exponents on the left and right sides of a level crossing
in Fig.\ 1(a) by $\gamma_{\rm L}$ and $\gamma_{\rm R}$, respectively,
we obtain a general relation of
\[
\gamma_{\rm L} \cdot \gamma_{\rm R}=
\frac{\Gamma_2}{\Gamma_1} \cdot \frac{\Gamma_1}{\Gamma_2}=1,
\]
since the roles of orbitals 1 and 2 are interchanged at the level crossing.

\section{Singlet-triplet Kondo effect}

We examine the Kondo effect around the singlet-triplet degeneracy for an
even number of electrons in this section.
In our previous papers,\cite{Eto,Eto2,Eto3}
the scaling analysis has been presented for this S-T Kondo effect in vertical
quantum dots in rectangular shape.\cite{Sasaki} Here, the analysis is
modified for circular quantum dots used in the experiment of ref.\ 24.

We consider a quantum dot with two orbitals and two electrons.
The exchange interaction which favors parallel spins,
\[
-E_{\rm ex}\sum_{s_1s_2s_3s_4}
({\mbi \sigma}/2)_{s_1s_2} \cdot ({\mbi \sigma}/2)_{s_3s_4}
d_{1,s_1}^{\dagger} d_{1,s_2} d_{2,s_3}^{\dagger} d_{2,s_4}
\]
with ${\mbi \sigma}=(\sigma_x,\sigma_y,\sigma_z)$,
is important in this problem. We add this term to the dot Hamiltonian
(\ref{eq:Hdot}) and fix ${\cal N}_{\rm{gate}} = 2$ for the
Coulomb blockade with two electrons in the quantum dot.
The possible ground state is a spin singlet ($S=0$) or triplet
($S=1$, $S_z \equiv M=1,0,-1$). We denote the states by $|S M \rangle$,
\begin{eqnarray*}
|0 0 \rangle & = &
d_{1 \uparrow}^{\dagger} d_{1 \downarrow}^{\dagger} |0 \rangle, \\
|1 1 \rangle & = &
d_{1 \uparrow}^{\dagger} d_{2 \uparrow}^{\dagger} |0 \rangle, \\
|1 0 \rangle & = & \frac{1}{\sqrt{2}}
(d_{1 \uparrow}^{\dagger} d_{2 \downarrow}^{\dagger}
+d_{1 \downarrow}^{\dagger} d_{2 \uparrow}^{\dagger}) |0 \rangle, \\
|1 -1 \rangle & = &
d_{1 \downarrow}^{\dagger} d_{2 \downarrow}^{\dagger} |0 \rangle,
\end{eqnarray*}
where $|0 \rangle$ represents the vacuum, assuming
$\varepsilon_1<\varepsilon_2$.\cite{com1}
The energy difference between them is $E_{S=0}-E_{S=1}=E_{\rm ex}
-(\varepsilon_2-\varepsilon_1)$, which is denoted by $\Delta$ in this
section. We disregard two other singlet states,
$d_{2 \uparrow}^{\dagger} d_{2 \downarrow}^{\dagger} |0 \rangle$ and
$(1/\sqrt{2})(d_{1 \uparrow}^{\dagger} d_{2 \downarrow}^{\dagger}
-d_{1 \downarrow}^{\dagger} d_{2 \uparrow}^{\dagger}) |0 \rangle$,
since they have larger energies than state $|0 0 \rangle$.
We assume that addition and extraction energies, $E^{\pm}$, are much
larger than $|\Delta|$ and level broadenings, $\Gamma_1$, $\Gamma_2$.

As in \S 2, we make an effective Hamiltonian $H_{\rm eff}$ in a
space of two-electron states, $|0 0 \rangle$, $|1 1 \rangle$,
$|1 0 \rangle$ and $|1 -1 \rangle$, taking account of the second-order
tunnel processes, $H_{\rm{T}}$. The Hamiltonian reads
\begin{equation}
H_{\rm{eff}}=H_{\rm{dot}}+H_{\rm{leads}}+
H^{S=1}+H^{S=1 \leftrightarrow 0}+H_{\rm{pot}}.
\end{equation}
The first term represents the dot state,
\begin{equation}
H_{\rm{dot}}=\sum_{S,M} E_{S} f_{SM}^{\dagger}f_{SM},
\end{equation}
using pseudofermion operators $f_{SM}^{\dagger}$ ($f_{SM}$) which
create (annihilate) the state $|SM \rangle$. It is required that
\begin{equation}
\sum_{SM} f_{SM}^{\dagger} f_{SM} =1.
\end{equation}
$H^{S=1}$ describes the spin-flip processes among three components of
the triplet state, $|1 1 \rangle$, $|1 0 \rangle$ and $|1 -1 \rangle$,
with a scattering of conduction electrons of channel 1 or 2.
The coupling constants are
given by $J_1=V_1^2/(2\tilde{E}_{\rm C})$ and $J_2=V_2^2/(2\tilde{E}_{\rm C})$,
respectively, where $1/\tilde{E}_{\rm C}=1/E^++1/E^-$.
$H^{S=1 \leftrightarrow 0}$ represents the conversion between the triplet
and singlet states with an interchannel scattering of conduction electrons.
The coupling constant is $\tilde{J}=V_1V_2/(\sqrt{2}\tilde{E}_{\rm C})$.
These spin-flip processes are schematically shown in Fig.\ 4.
A potential scattering $H_{\rm{pot}}$ with
$T=(V_1^2+V_2^2)/(2\tilde{E}_{\rm C})$ is also relevant to the S-T Kondo
effect. (See ref.\ 19 
for explicit expressions of the Hamiltonian.)

\begin{figure}[t]
\begin{center}
\includegraphics[width=6cm]{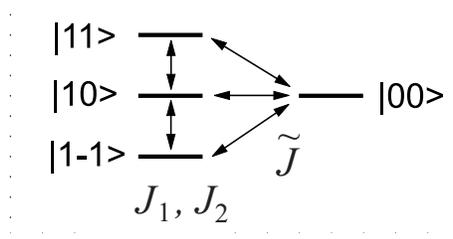}
\end{center}
\caption{Four states that we consider in a quantum dot with two orbitals
and two electrons, $| S,M \rangle$,
and exchange processes among them.
The exchange process denoted by $J_1$ ($J_2$) is accompanied by a spin-flip
scattering of conduction electrons of channel 1 (2), whereas that by
$\tilde{J}$ is accompanied by an interchannel scattering of conduction
electrons.}
\end{figure}

We obtain the scaling equations for the coupling constants,
$J_1$, $J_2$, $\tilde{J}$, $T$, using the poor man's scaling method.
When $D \gg |\Delta|$,
\begin{equation}
\left\{
\begin{array}{rcl}
d J_1 / \nu d \ln D & = & -2 J_1^2 - \tilde{J}^2, \\
d J_2 / \nu d \ln D & = & -2 J_2^2 - \tilde{J}^2, \\
d \tilde{J}/ \nu d \ln D & = & -2 (J_1+J_2) \tilde{J} - T \tilde{J}, \\
d T/ \nu d \ln D & = &  -8 \tilde{J}^2.
\end{array}
\right.
\label{eq:scalA}
\end{equation}
These equations have a stable fixed point of
$J_1=J_2=\infty$,
$\tilde{J}/J_1=\sqrt{2(2+\sqrt{5})}$ and $T/J_1=2(1+\sqrt{5})$
(Appendix B).
The renormalization flow of the exchange couplings goes toward this fixed
point, as the energy scale $D$ decreases.

When $D \ll \Delta$, the singlet state can be safely disregarded.
The scaling equations are given by
\begin{equation}
\left\{
\begin{array}{rcl}
d J_1 / d \ln D & = & -2 \nu J_1^2, \\
d J_2 / d \ln D & = & -2 \nu J_2^2,
\end{array}
\right.
\label{eq:scalB}
\end{equation}
indicating that $J_1$ and $J_2$ evolve independently. The other coupling
constants do not change.
On the singlet side ($\Delta<0$), all the exchange couplings
saturate, provided $D \ll |\Delta|$.

The Kondo temperature $T_{\rm K}$ is evaluated as a function of $\Delta$,
in the same way as in \S 3. (i) When $|\Delta| \ll T_{\rm K}(0)$,
the scaling equations (\ref{eq:scalA}) are valid till the end of the scaling.
Then $T_{\rm K}$ is maximal although the analytical expression of
$T_{\rm K}(0)$ cannot be obtained.\cite{com2}

(ii) When $\Delta > D_0$, the ground state is a spin triplet, whereas the
singlet state is not relevant to the Kondo effect.
Then the scaling equations (\ref{eq:scalB}) work in the whole scaling region.
This yields
\begin{equation}
T_{\rm K}(\infty) = D_0 \exp [-1/2\nu J_1],
\label{eq:TK0b}
\end{equation}
when $J_1 \ge J_2$ ($J_1$ should be replaced by $J_2$ when $J_1 < J_2$).
This is the Kondo temperature of a localized
spin with $S=1$.\cite{Okada}

(iii) In the intermediate region of $T_{\rm K}(0) \ll \Delta \ll D_0$,
all the coupling constants develop with decreasing $D$, following
eq.\ (\ref{eq:scalA}) when $D \gg \Delta$.
$\tilde{J}$ and $T$ saturate at $D \approx \Delta$,
whereas $J_1$ and $J_2$ continue to grow by eq.\  (\ref{eq:scalB}) when
$D \ll \Delta$. To see the $\Delta$ dependence of $T_{\rm K}$,
we perform numerical calculations.\cite{Eto3}
We solve eqs.\ (\ref{eq:scalA}) and (\ref{eq:scalB}) for $D>\Delta$ and
$D<\Delta$, respectively, with the initial condition of
$\nu (J_1+J_2) = 0.01$. The Kondo temperature $T_{\rm K}$ is determined
as the energy scale $D$ at which $\nu J_1 = 1$ or $\nu J_2 = 1$.
The calculated results of
$T_{\rm K}(\Delta)$ are shown in Fig.\ 5 on a log-log scale, with
(a) $J_2/J_1=(V_2/V_1)^2=1$, (b) $3/4$ or $4/3$, and (c) $1/2$ or $2$.
$T_{\rm K}(\Delta)$ shows the power law of eq.\ (\ref{eq:powerDD2})
with an exponent of $\gamma=2+\sqrt{5}$ if $\Delta/T_{\rm K}(0)$ is
not too large (Appendix B).\cite{Pustilnik,Pustilnik2}
With increasing $\Delta$, the exponent gradually deviates from the universal
value to a nonuniversal one,
Min$\{ \Gamma_2/\Gamma_1, \Gamma_1/\Gamma_2 \}$.\cite{Eto,Eto2,Eto3}

(iv) On the singlet side ($\Delta<0$), no Kondo effect is expected when
$|\Delta| \gg T_{\rm K}(0)$. $T_{\rm K}(\Delta)$ drops to zero
suddenly at $|\Delta| \sim T_{\rm K}(0)$.
The behavior of $T_{\rm K}(\Delta)$ is schematically shown in Fig.\ 6(b).

\begin{figure}[t]
\begin{center}
\includegraphics[width=6cm]{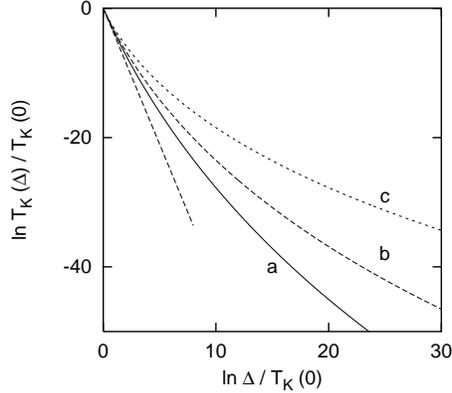}
\end{center}
\caption{
Calculated results of the Kondo temperature $T_{\rm K}$ for the S-T Kondo
effect, as a function of the energy difference $\Delta$ between spin-singlet
and -triplet states, on a log-log scale.
Both $T_{\rm K}$ and $\Delta$ are normalized by the Kondo temperature at
$\Delta=0$, $T_{\rm K}(0)$.
(a) $(V_2/V_1)^2=1$, (b) $3/4$ or $4/3$, and (c) $1/2$ or $2$.
The scaling equations, (\ref{eq:scalA}) for $D>\Delta$ and
(\ref{eq:scalB}) for $D<\Delta$, are solved numerically
with the initial condition of $\nu(J_1+J_2) = 0.01$.
A straight broken line
indicates a slope of $\gamma=2+\sqrt{5} \approx 4.2$.}
\end{figure}

\section{Conclusions}

We have theoretically examined D-D and S-T Kondo effects in multilevel
quantum dots. For the D-D Kondo effect, we have considered
a quantum dot with two orbitals and spin $1/2$.
Using the poor man's scaling method, the Kondo
temperature $T_{\rm K}$ has been evaluated as a function of energy separation
$\Delta$ between the two orbitals.
When the tunnel couplings are equivalent for the two orbitals,
$V_1=V_2$ in eq.\ (\ref{eq:tunnel}),
$T_{\rm K}(\Delta)$ is maximal around $\Delta = 0$, and
decreases with increasing $|\Delta|$ obeying a power law,
eq.\ (\ref{eq:powerDD}), with exponent $\gamma=1$. This behavior of
$T_{\rm K}(\Delta)$ is understood as a crossover
from SU(4) to SU(2) Kondo effect. In general case with different tunnel
couplings for the two orbitals, $V_1 \ne V_2$,
we observe a power law with nonuniversal
exponents, reflecting the marginality of the fixed point of SU(4) Kondo effect.
We find a relation of $\gamma_{\rm L} \cdot \gamma_{\rm R}=1$,
where $\gamma_{\rm L}$ and $\gamma_{\rm R}$ are the exponents on both sides
of a level crossing, as long as $V_1 \sim V_2$.

\begin{figure}[t]
\begin{center}
\includegraphics[width=6cm]{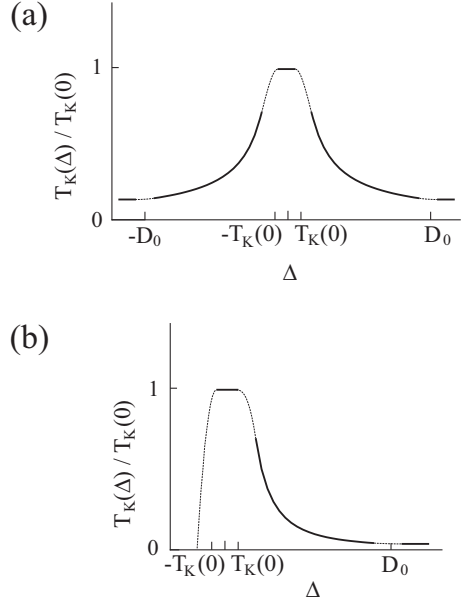}
\end{center}
\caption{Schematical drawing of the Kondo temperature $T_{\rm K}$, as
a function of $\Delta$, for (a) D-D Kondo effect and (b) S-T Kondo effect.
$\Delta=\varepsilon_2-\varepsilon_1$ is the level spacing between
two orbitals in (a), whereas
$\Delta=E_{S=0}-E_{S=1}$ is the energy difference between spin-singlet
and -triplet states in (b).
The tunnel couplings are equivalent for two orbitals; $V_1 = V_2$
in eq.\ (\ref{eq:tunnel}). $T_{\rm K}$ is normalized by
the Kondo temperature at $\Delta=0$, $T_{\rm K}(0)$. $D_0$ is the bandwidth
of conduction electrons in the leads.}
\end{figure}

For the S-T Kondo effect, we have examined a quantum dot with two orbitals and
two electrons. The Kondo temperature is maximal around $\Delta = 0$,
where $\Delta$ is the energy difference between spin-singlet and -triplet
states. $T_{\rm K}(\Delta)$ decreases with increasing
$|\Delta|$ in a different way on the triplet side ($\Delta>0$) or
on the singlet side ($\Delta<0$): $T_{\rm K}(\Delta)$  approximately
shows a power law with $\gamma=2+\sqrt{5}$ when $\Delta>0$,
whereas $T_{\rm K}$ drops to zero suddenly at $|\Delta| \sim T_{\rm K}$
when $\Delta<0$.

In both D-D and S-T Kondo effects, the Kondo temperature is maximal
around the degeneracy point ($\Delta=0$), indicating that a total of fourfold
spin and orbital degeneracy contributes to the enhancement of the Kondo
effect. Now we compare $T_{\rm K}(\Delta)$ of these Kondo effects in details to
explain recent experimental results.\cite{Sasaki2}
In the experimental situation of $V_1 \approx V_2$,
there are two major differences between them (see Fig.\ 6).

First, when the energy difference $\Delta$ is tuned by a magnetic field,
the behavior of $T_{\rm K}(\Delta)$ is almost symmetric in D-D case and
largely asymmetric in S-T case, with respect to the degeneracy point
($\Delta=0$) in Fig.\ 1. In the former, $T_{\rm K}$ decreases gradually with
increasing $|\Delta|$ obeying a power law of eq.\ (\ref{eq:powerDD}) on
both sides of $\Delta>0$ and $\Delta<0$.
In the latter, $T_{\rm K}(\Delta)$ approximately obeys a power law
on the triplet side ($\Delta>0$), whereas it drops to zero abruptly on
the singlet side ($\Delta<0$).

Second, the exponent of eq.\ (\ref{eq:powerDD}) is about unity in D-D Kondo
effect when $V_1 \approx V_2$. The value is smaller than $\gamma=2+\sqrt{5}$,
the exponent in S-T Kondo effect on the triplet side. Therefore, $T_{\rm K}$
decreases with increasing $|\Delta|$ more slowly in D-D case.

In ref.\ 24, 
both D-D and S-T Kondo effects
have been observed in the same quantum dot with different gate voltage
and magnetic field. The enhancement of conductance due to the Kondo effect,
$\Delta G$, has been examined in the vicinity of D-D and S-T degeneracy
points. As the magnetic field is changed from the degeneracy point,
$\Delta G$ drops asymmetrically in the S-T case, whereas $\Delta G$ decreases
more slowly and symmetrically in the D-D case. These behaviors of $\Delta G$
are in accordance with our theoretical results of $T_{\rm K}(\Delta)$
although the direct measurement of $T_{\rm K}(\Delta)$ has not been reported.

Finally, we comment on the electronic states at $T \ll T_{\rm K}$.
In the D-D Kondo effect, the dot state is represented by a pair of orbital
index ($i=1,2$) and spin ($S_z=\pm 1/2$). They are fully screened by the
electrons in the leads with two conduction channels if
$|\Delta| \ll T_{\rm K}$. In consequence an usual Fermi liquid state of SU(4)
symmetry appears in which the orbital and spin degrees of freedom are
totally entangled, as discussed by Borda {\it et al}.\cite{Borda}
The Fermi liquid properties are not changed by the marginality of
the SU(4) Kondo effect.
If $|\Delta| \gg T_{\rm K}$, the upper orbital is not relevant to the
Kondo state. A single channel screens an electron spin in the lower
orbital, whereas the other channel is decoupled. Then we should observe
a usual Fermi liquid state. Despite the different properties,
the conductance through the quantum dot is
$G=2e^2/h$ [apart from an asymmetric factor of two tunnel junctions]
in both cases.\cite{Sakano}
(i) When $|\Delta| \ll T_{\rm K}$, two conduction channels contribute to the
transport, but each channel has a phase shift of $\pi/4$ in the SU(4) Kondo
effect.\cite{Borda,Sakano} Hence the total conductance is $2e^2/h$.
(ii) When $|\Delta| \gg T_{\rm K}$, a single channel takes part in the
transport. The phase shift of $\pi/2$ in the conventional SU(2) Kondo effect
(unitary limit) results in $G=2e^2/h$.

In the S-T Kondo effect, the spin-triplet state is fully screened by
two conduction channels. The phase shift of each channel is given by
$\pi/2$ in this case. Thus we observe $G=4e^2/h$ everywhere on the triplet
side at $T=0$.
The $T$ dependence of the conductance has been discussed in a special
case of $V_1=V_2$ by Pustilnik {\it et al}.\cite{Pustilnik,Pustilnik2}

\acknowledgements

The author gratefully acknowledges discussions with Yu.\ V.\ Nazarov,
L.\ I.\ Glazman, H.\ Hyuga, N.\ Kawakami, S.\ Sasaki and S.\ Tarucha.

\appendix
\section{Fixed point of D-D Kondo effect}

For the D-D Kondo effect, the scaling equations for the exchange couplings,
$J_1$, $J_2$, $\tilde{J}$, $\tilde{T}$ and $T$,
are given by eq.\ (\ref{eq:DDscale}) for $D \gg \Delta$.
To find and analyze the fixed point, it is convenient to consider the
equations for $j_{\rm A}=J_{\rm A}/J_{\rm S}$,
$\tilde{j}=\tilde{J}/J_{\rm S}$,
$\tilde{t}=\tilde{T}/J_{\rm S}$ and $t=T/J_{\rm S}$, where
$J_{\rm S}=J_1+J_2$ and $J_{\rm A}=J_1-J_2$. They are given by
\begin{eqnarray}
\frac{d j_{\rm A}}{d x} & = &
\frac{2j_{\rm A}}{1+j_{\rm A}^2+2\tilde{j}(\tilde{j}+\tilde{t})}-j_{\rm A},
\label{eq:DDeq1} \\
\frac{d \tilde{j}}{d x} & = &
\frac{\tilde{j}+\tilde{t}/2+\tilde{j}t}
{1+j_{\rm A}^2+2\tilde{j}(\tilde{j}+\tilde{t})}-\tilde{j},
\label{eq:DDeq2} \\
\frac{d \tilde{t}}{d x} & = &
\frac{3\tilde{j}/2+\tilde{t}/2+\tilde{j}t}
{1+j_{\rm A}^2+2\tilde{j}(\tilde{j}+\tilde{t})}-\tilde{t},
\label{eq:DDeq3} \\
\frac{d t}{d x} & = &
\frac{3\tilde{j}^2+\tilde{t}^2}
{1+j_{\rm A}^2+2\tilde{j}(\tilde{j}+\tilde{t})}-t,
\label{eq:DDeq4}
\end{eqnarray}
with $x=\ln (\nu J_{\rm S})$ (we assume that $J_{\rm S}$ goes to infinity
at the fixed point). $J_{\rm S}$ obeys the scaling equation
\begin{equation}
\frac{d}{d {\cal L}}(1/\nu J_{\rm S})=
1+j_{\rm A}^2+2\tilde{j}(\tilde{j}+\tilde{t}),
\label{eq:DDeq0}
\end{equation}
where ${\cal L}=\ln D/D_0$.

From scaling equations, (\ref{eq:DDeq1})--(\ref{eq:DDeq4}),
we find a fixed point of
\begin{equation}
j_{\rm A}=0, \ \tilde{j}=\tilde{t}=t=1/2,
\end{equation}
with $J_{\rm S}=\infty$, which corresponds to the fixed point of
SU(4) Kondo effect.
By the linearization of the scaling equations around the fixed point,
we find that the fixed point is marginal. A marginal valuable is $j_{\rm A}$.

We derive the power law of $T_{\rm K}(\Delta)$, eqs.\ (\ref{eq:powerDD2}) and
(\ref{eq:DDgamma}), approximately in the following.
First, we expand $1/\nu J_{\rm S}(D)$ around $D=T_{\rm K}(0)$
(${\cal L}=\ln T_{\rm K}(0)/D_0 \equiv {\cal L}_{\rm K}$) at which
$1/\nu J_{\rm S}=0$. We assume that $\tilde{j}=\tilde{t}=t=1/2$
(values at the fixed point), whereas the marginal variable $j_{\rm A}$ is
fixed at the initial value, $(V_1^2-V_2^2)/(V_1^2+V_2^2)$, at $D=T_{\rm K}(0)$.
To the first order in ${\cal L}-{\cal L}_{\rm K}=\ln D/T_{\rm K}(0)$,
\begin{eqnarray}
1/\nu J_{\rm S}(D) & \approx & 
\left. \frac{d}{d {\cal L}}(1/\nu J_{\rm S}) \right|_{{\cal L}_{\rm K}}
({\cal L}-{\cal L}_{\rm K}) \nonumber \\
& = & (2+j_{\rm A}^2)({\cal L}-{\cal L}_{\rm K}).
\label{eq:DDtmp}
\end{eqnarray}
For $T_{\rm K}(0) \ll |\Delta| \ll D_0$, $T_{\rm K}(\Delta)$ is determined
as the energy scale at which $1/\nu J_1=0$ in the solution of
eq.\ (\ref{eq:DDscale2}),
\begin{equation}
1/\nu J_1(D)=-2 \ln T_{\rm K}(\Delta)/D,
\label{eq:DDtmp2}
\end{equation}
Connecting eq.\ (\ref{eq:DDtmp}) [$J_1(D)=J_{\rm S}(D)(1+j_{\rm A})/2$] and
eq.\ (\ref{eq:DDtmp2}) at $D=\Delta$, we obtain
the power law of eq.\ (\ref{eq:powerDD2}) with
\begin{equation}
\gamma=\frac{1-j_{\rm A}+j_{\rm A}^2}{1+j_{\rm A}}
=\left( V_2/V_1 \right)^2+O(j_{\rm A}^2).
\end{equation}
This is identical to eq.\ (\ref{eq:DDgamma}) if $j_{\rm A}^2$ is neglected.

Apart from the fixed point (with large $\Delta$),
another approximation is taken to explain the power law of $T_{\rm K}(\Delta)$.
We expand $1/\nu J_{\rm S}$ around the initial point ($D=D_0$, ${\cal L}=0$).
By the substitution of the initial values,
$j_{\rm A}=(V_1^2-V_2^2)/(V_1^2+V_2^2)$,
$\tilde{j}=V_1V_2/(V_1^2+V_2^2)$ and $\tilde{t}=1/2$,
into the right-hand side of eq.\ (\ref{eq:DDeq0}), we obtain
\begin{equation}
1/\nu J_{\rm S}(D) - 1/\nu J_{\rm S}(D_0) \approx 2 {\cal L}.
\label{eq:DDtmp3}
\end{equation}
If we connect eqs.\ (\ref{eq:DDtmp3}) and (\ref{eq:DDtmp2}) at $D=\Delta$,
we find $T_{\rm K}(\Delta) \propto \Delta^{-\gamma}$ with exponent of
eq.\ (\ref{eq:DDgamma}).
This implies that the power law of eqs.\ (\ref{eq:powerDD2}) and
(\ref{eq:DDgamma}) holds in a wide range of $\Delta$, which is
consistent with numerical results shown in Fig.\ 3.

\section{Fixed point of S-T Kondo effect}

For the S-T Kondo effect, the scaling equations (\ref{eq:scalA}) for
$D \gg \Delta$ are analyzed in the same way as in Appendix A.
The scaling equations for $j_{\rm A}=J_{\rm A}/J_{\rm S}$,
$\tilde{j}=\tilde{J}/J_{\rm S}$ and $t=T/J_{\rm S}$,
where $J_{\rm S/A}=J_1 \pm J_2$, have a fixed point of
\begin{equation}
j_{\rm A}=0, \ \tilde{j}=\sqrt{(2+\sqrt{5})/2}, \ t=1+\sqrt{5},
\label{eq:STfixp}
\end{equation}
with $J_{\rm S}=\infty$. This is a stable fixed point.

The scaling equation for $J_{\rm S}$ is
\begin{equation}
\frac{d}{d {\cal L}}(1/\nu J_{\rm S})=1+j_{\rm A}^2+2\tilde{j}^2,
\label{eq:eqST0}
\end{equation}
where ${\cal L}=\ln D/D_0$. We expand the solution of eq.\ (\ref{eq:eqST0})
around the fixed point of eq.\ (\ref{eq:STfixp}) at which
$D=T_{\rm K}(0)$ (${\cal L}={\cal L}_{\rm K}$).
To the first order in ${\cal L}-{\cal L}_{\rm K}=\ln D/T_{\rm K}(0)$,
\begin{equation}
1/\nu J_{\rm S}(D) \approx (3+\sqrt{5})({\cal L}-{\cal L}_{\rm K}).
\label{eq:STtmp}
\end{equation}
For $T_{\rm K}(0) \ll |\Delta| \ll D_0$,
we match eq.\ (\ref{eq:STtmp}) [$J_1(D)=J_2(D)=J_{\rm S}(D)/2$] and
the solution of eqs.\ (\ref{eq:scalB}) at $D=\Delta$,\cite{Pustilnik}
and find a power law, eq.\ (\ref{eq:powerDD2}), with
\begin{equation}
\gamma=2+\sqrt{5}.
\end{equation}

The universal exponent of the power law is observed in a limited range of
$\Delta$ in Fig.\ 5. Far away from the fixed point, we expand $1/\nu J_{\rm S}$
around the initial point of $D=D_0$ (${\cal L}=0$).
Substituting the initial values, $j_{\rm A}=(V_1^2-V_2^2)/(V_1^2+V_2^2)$
and $\tilde{j}=\sqrt{2}V_1V_2/(V_1^2+V_2^2)$,
into the right-hand side of eq.\ (\ref{eq:eqST0}), we obtain
\begin{equation}
1/\nu J_{\rm S}(D) - 1/\nu J_{\rm S}(D_0) \approx 2 {\cal L}.
\label{eq:STtmp2}
\end{equation}
We match eq.\ (\ref{eq:STtmp2}) [$J_{1,2}(D)=J_{\rm S}(D)(1 \pm j_{\rm A})/2$]
and the solution of eqs.\ (\ref{eq:scalB}) at $D=\Delta$, we find
a power law, $T_{\rm K}(\Delta) \propto \Delta^{-\gamma}$, with exponent of
\begin{equation}
\gamma=\left( \frac{V_2}{V_1} \right)^2=\frac{\Gamma_2}{\Gamma_1},
\label{eq:STgamma}
\end{equation}
if $\Gamma_1 \ge \Gamma_2$ ($J_1 \ge J_2$ and hence $J_1$ diverges faster
than $J_2$), whereas
\begin{equation}
\gamma=\left( \frac{V_1}{V_2} \right)^2=\frac{\Gamma_1}{\Gamma_2},
\label{eq:STgamma2}
\end{equation}
if $\Gamma_1 < \Gamma_2$. Note that $0<\gamma \le 1$ in this case.
In consequence,
the exponent of the power law of $T_{\rm K}(\Delta)$ changes gradually
from the universal value, $2+\sqrt{5}$, to
a nonuniversal one given by eq.\ (\ref{eq:STgamma}) or (\ref{eq:STgamma2}),
with increasing $\Delta$, as shown in Fig.\ 5.\cite{Eto3}
This is in contrast to the case of Fig.\ 3 for the D-D Kondo effect in
which a power law with fixed exponent holds in a wide range of $\Delta$.


\begin{thebibliography}{99}
\bibitem{com1}
In our previous papers,\cite{Eto,Eto2,Eto3}
we have considered a singlet state of
$|0 0 \rangle =
(C_1 d_{1 \uparrow}^{\dagger} d_{1 \downarrow}^{\dagger}
-C_2 d_{2 \uparrow}^{\dagger} d_{2 \downarrow}^{\dagger}) |0 \rangle$,
where $|C_1|^2+|C_2|^2=1$. In general, $C_2 \ne 0$ owing to
the matrix element of the Coulomb interaction between
$d_{1 \uparrow}^{\dagger} d_{1 \downarrow}^{\dagger} |0 \rangle$
and $d_{2 \uparrow}^{\dagger} d_{2 \downarrow}^{\dagger} |0 \rangle$,
$\langle 22|e^2/r|11 \rangle$. 
In this paper, we fix $C_1=1$ and $C_2=0$ to examine the
experimental situation of circular dots,\cite{Sasaki2} in which
the total angular momentum conserves and hence the matrix element is absent
if orbitals 1 and 2 have different angular momentums.
\bibitem{com2}
A lower limit of $T_{\rm K}(0)$ is given by
$D_0 \exp [-1/2\nu (J_1+J_2)]$.\cite{Eto3}
\end{thebibliography}
\end{document}